\pdfoutput=1
\documentclass[aps,prl,reprint,groupedaddress,amsmath,amssymb]{revtex4-2}
\usepackage{graphicx}% Include figure files
\usepackage{bm}% bold math

\usepackage{placeins} % for \FloatBarrier
\usepackage{longtable}

\begin{document}

%\preprint{}

\title{MESA Technical Note: Beam Breakup Instability Threshold Current}


\author{S.~Glukhov}
\email[]{sergei.glukhov@tu-darmstadt.de}
%\thanks{}
\author{O.~Boine-Frankenheim}
\affiliation{TEMF TU Darmstadt, Darmstadt}

\author{C.~Stoll}
\author{F.~Hug}
\affiliation{Institut f{\"u}r Kernphysik, JGU Mainz}


\date{\today}

\begin{abstract}
MESA (Mainz Energy recovery Superconducting Accelerator) is an energy recovery linac (ERL) which is under construction at Johannes Gutenberg University in Mainz. It will be operated in external beam (EB) mode with 150~$\mu$A electron beam at 155~MeV and energy recovery (ER) mode with 1~mA (first stage) and later 10~mA (second stage) electron beam at 105~MeV. An important factor which may limit performance of the machine is a beam breakup (BBU) instability which may occur due to excitation of higher-order modes (HOMs) in superconducting RF cavities. This effect occurs only when the injected beam current exceeds a threshold value. The aim of the present work is to develop a software for reliable determination of the threshold current in MESA, find main factors which may change its value and finally make a decision concerning capability of MESA operation at 10~mA and need for additional measures for suppressing BBU instability.
\end{abstract}

%\keywords{}

\maketitle

\section{Introduction}

\begin{figure}[h!]
	\includegraphics[width=86mm]{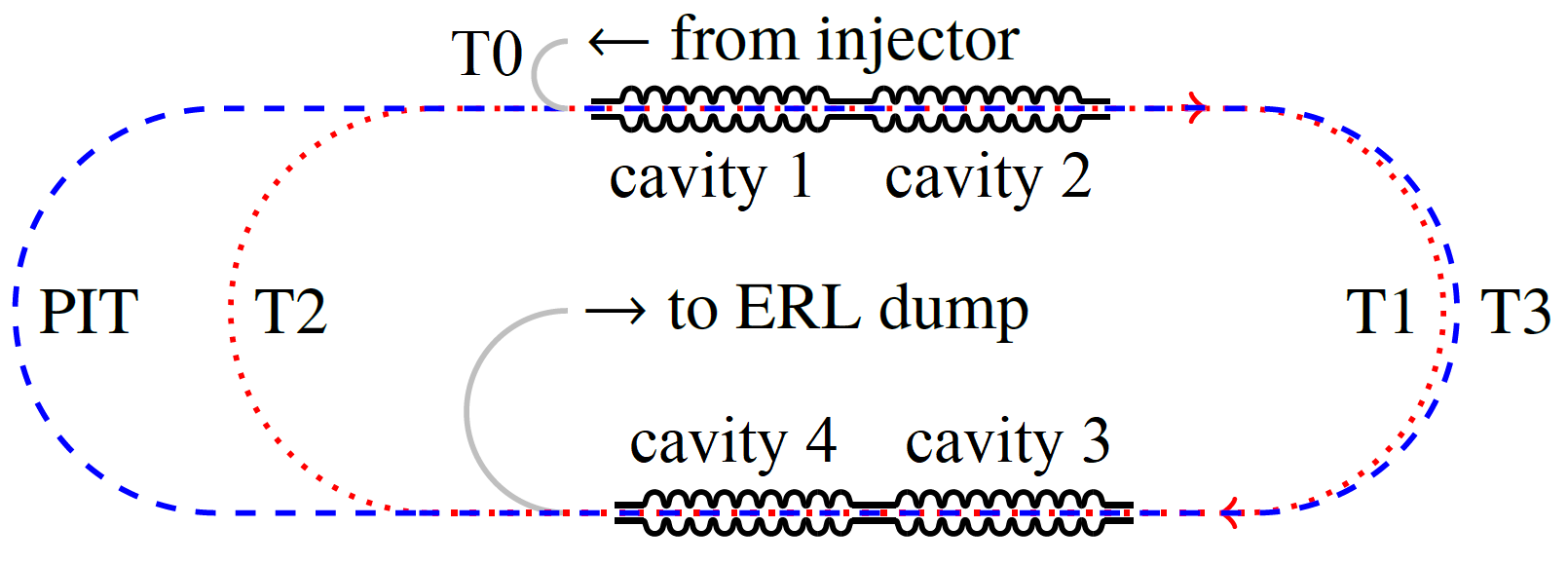}
	\caption{\label{fig:MESA}Schematic view of MESA.}
\end{figure}

\begin{figure}[h!]
	\includegraphics[width=86mm]{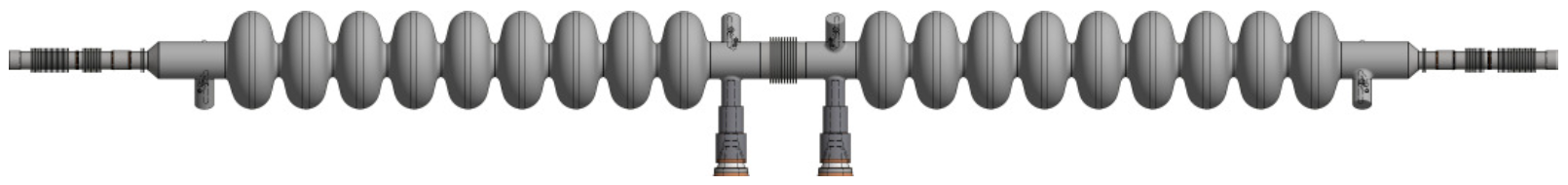}
	\caption{\label{fig:AccelModule}CAD model of MESA accelerating module.}
\end{figure}

The schematic MESA layout in ER mode is presented in Fig.~\ref{fig:MESA}. The machine consists of the injection arc T0, two accelerating modules connected with recirculation arcs T1, T2, T3 and a pseudo internal target (PIT) arc where experiments are performed. Each accelerating module consists of two 9\nobreakdash-cell TESLA cavities connected as shown in Fig.~\ref{fig:AccelModule}. 5~MeV electron bunches are injected through T0 and move through T1, T2 and T3 gaining 12.5~MeV on each cavity pass. The length of PIT arc is adjusted in a way that bunches arrive to the next cavity with a $180^{\circ}$ phase shift (in decelerating phase) and then move through T3, T2 and T1 losing 12.5~MeV on each cavity pass \cite{Hug:2017ypc}.

The BBU instability mechanism is as follows. Electron bunches periodically injected into ERL excite parasitic HOMs in superconducting RF cavities. These HOMs may act back on the bunches on subsequent passes. Dipole modes have the most impact because among other non-axisymmetric modes their field is concentrated closer to the cavity axis. A bunch kicked by a HOM receives an additional transversal momentum. When the bunch returns into the same RF cavity after recirculation this momentum transforms into a transversal offset. A transversally shifted bunch changes HOM voltage in the cavity, the change depends on the HOM voltage value of the previous recirculation. This may lead to unstable motion \cite{Stoll:2018dij, Stoll:2019qtg}.

\section{Main Definitions}

For the purpose of BBU simulations an accelerator can be represented as a sequence of arcs and thin RF cavities at the end of each arc in the order seen by one bunch on its way from injector to beam dump. Each cavity may appear several times in this sequence depending on number of recirculations. Regions between two cavities of the same accelerating module are straight but for simplicity they will be also referred to as arcs in the following text. Each cavity is assigned with a certain number of dipole HOMs. Adjacent HOMs form a pair with close frequencies and nearly orthogonal polarization. In all subsequent simulations the same (and only one) HOM pair is excited in all cavities. The beam is represented as a sequence of pointlike on-energy equidistantly injected bunches. It is necessary to clearly define the characteristics for each of these entities:
\begin{enumerate}
	\item Accelerator:
	\begin{itemize}
		\item accelerated particle rest energy $E_0$;
		\item RF voltage period $t_{\textrm{RF}}$;
		\item number of cavities $N_c$;
		\item number of HOMs $N_m$ in each cavity\\($N_m = 2$ in all simulations);
		\item number of arcs $N_a$.
	\end{itemize}
	\item Arc:
	\begin{itemize}
		\item $6 \times 6$ transport matrix $\mathbf{T}^{(i)}$;
		\item length $L_i$;
		\item reference particle energy $E_i$;
		\item reference particle velocity\\$v_i = c \sqrt{1 - \left( E_0 / E_i \right)^2}$;
		\item reference particle time of flight $\Delta t_i = L_i / v_i$.
	\end{itemize}
	\item Dipole HOM:
	\begin{itemize}
		\item frequency $f_{\lambda}$ or $\omega_{\lambda} = 2 \pi f_{\lambda}$;
		\item Q-factor $Q_{\lambda}$;
		\item damping factor $\Gamma_{\lambda} = \frac{\omega_{\lambda}}{2 Q_{\lambda}}$;
		\item shunt impedance $(R/Q)_{\lambda}$;
		\item reduced shunt impedance \\ $(R/Q)'_{\lambda} = (R/Q)_{\lambda} (\omega_{\lambda}/c)^2$;
		\item polarization angle $\theta_{\lambda}$.
	\end{itemize}
	\item Beam:
	\begin{itemize}
		\item injector current $I_0$;
		\item number of injected bunches $N_b$;
		\item time interval between bunches $t_b$ \\ ($t_b = t_{\textrm{RF}}$ for the MESA case);
		\item bunch charge $q = I_0 t_b$.
	\end{itemize}
	\item Bunch:
	\begin{itemize}
		\item 6-component accelerator coordinate vector (relative to reference particle) $\mathbf{X} = \left( x, p_x, y, p_y, z, p_z \right)^T$;
		\item longitudinal coordinate $Z$ along reference orbit in laboratory frame.
	\end{itemize}
	\item Combined entities:
	\begin{itemize}
		\item average particle momentum in the cavity at the end of the $i$-th arc\\$p_i = (E_i v_i / c^2 + E_{i+1} v_{i+1} / c^2)/2$;
		\item HOM $\lambda$ voltage in the $j$-th cavity $V_{\lambda}^{(j)}$.
	\end{itemize}
\end{enumerate}

A small remark concerning HOM properties is necessary. An on-axis particle can be kicked by a HOM transversally; a line perpendicular to the direction of the kick is called polarization axis. The polarization angle is an angle between the horizontal axis (in accelerator coordinates) and the normal to the polarization axis. The prime in $(R/Q)'_{\lambda}$ (measured in $\Omega/\textrm{m}^2$) is introduced to distinguish it from $(R/Q)_{\lambda}$ (measured in $\Omega$) used in \cite{Krafft:1990BBU} and \cite{Hoffstaetter:2004BBU}. The usage of $(R/Q)'_{\lambda}$ leads to simpler formulae because the impedance of the dipole HOMs is quadratically dependent on the distance from the polarization axis and this is the proportionality coefficient.

\section{Bunch Tracking}

When a bunch passes the $i$\nobreakdash-th arc followed by the $j$\nobreakdash-th cavity, its coordinates are transformed as follows:
\[
	\mathbf{X} \mapsto \mathbf{T}^{(i)} \mathbf{X} + \frac{1}{p_i} \sum_{\lambda} \Im V_{\lambda}^{(j)} \left( 0, \cos \theta_{\lambda}, 0, \sin \theta_{\lambda}, 0, 0 \right)^T \,.
\]
The HOM voltages in the cavity also change:
\begin{eqnarray}
	\label{eq:Vtransformation}
	V_{\lambda}^{(j)} \!\mapsto\! V_{\lambda}^{(j)} e^{\left(i \omega_{\lambda} - \Gamma_{\lambda}\right) \Delta t_i} \!+\! \frac{q}{2} \left(\frac{R}{Q}\right)'_{\!\lambda} \!\!(x \cos \theta_{\lambda} + y \sin \theta_{\lambda}) \,, \nonumber \\
	\lambda = 1 \dots N_m \,. \nonumber \\
\end{eqnarray}

The most obvious tracking algorithm is to unroll the lattice onto the positive semiaxis, place initially all the bunches on negative semiaxis equidistantly and then move them through the lattice with their velocities stopping each time one of them encounters a cavity. Usually in simulations $N_b > 10^5$, therefore it is important to use a tracking algorithm which has complexity not more than linear by this parameter. On each step a bunch should be found which is the first to encounter a cavity, but not all the bunches should be checked, only the first bunch in each arc. When a bunch leaves an arc, the next one becomes the first. Therefore number of bunches to check is always not more than $N_a$.

The longitudinal position of all bunches also should not be updated on each step. It is enough to have a laboratory time counter and an individual counter for each bunch which shows when its position was updated last time. A similar counter is also necessary for each cavity to show when HOM voltages in it were updated last time. The difference between this counter and laboratory time should be used in (\ref{eq:Vtransformation}) instead of $\Delta t_i$.

\section{Beam Breakup Instability Theory (Simplest Case)}

The main equations of the BBU theory were derived in \cite{Krafft:1990BBU} and \cite{Hoffstaetter:2004BBU}. The approach developed in \cite{Krafft:1990BBU} is much simpler but suitable only for the case of recirculating linac where initially injected and all the recirculated bunches are at the same RF phase. Laplace transformation of $V_{\lambda}(t)$ is introduced in \cite{Hoffstaetter:2004BBU} to overcome this limitation. It will be shown in this section that such a complication is not necessary and the approach used in \cite{Krafft:1990BBU} can be easily extended to the case of an arbitrary RF phase of the recirculated bunches.

Consider a charge $q_e$ moving at the speed of light $c$ through an accelerator parallel to its axis with a transverse offset $d$ and a test particle moving on-axis with the same speed and time delay $t$. Transversal wake function $W_t(t)$ is a transversal Lorentz force acting on the test particle integrated over path of interest per unit charges and transversal offset:
\[
	W_t(t) = \frac{1}{q_e d} \int_{\textrm{path}} \left( E_x(z,z/c+t) - c B_y(z,z/c+t) \right) dz \,.
\]
For point charges in an RF cavity this becomes:
\begin{equation}
	\label{eq:Wt_t}
	W_t(t) =
	\begin{cases}
		\displaystyle \sum_{\lambda} \left(\frac{R}{Q}\right)_{\lambda} \frac{\omega_{\lambda}{}^2}{2 c} e^{-\Gamma_{\lambda} t} \sin (\omega_{\lambda} t), & t \geqslant 0 \\
		0, & t < 0
	\end{cases} \,,
\end{equation}
where index $\lambda$ iterates over all HOMs which can be excited in the cavity.

Consider an ERL with one recirculation turn and one thin RF cavity containing one dipole HOM. The HOM voltage $V_{\lambda}(t)$ had given additional deflection to the bunch; this deflection transformed into additional transversal offset $d(t)$ after one recirculation:
\begin{equation}
	\label{eq:d_t}
	d(t) = T_{12} \frac{q_e}{p c} V_{\lambda}(t - t_r) \,,
\end{equation}
where $t_r$ is recirculation time. On the other hand, a beam current $I(t)$ passing the cavity at a transversal position $d(t)$ creates the following deflecting voltage:
\begin{equation}
	\label{eq:V_int}
	V_{\lambda}(t) = \int_{-\infty}^{t} W_t(t - t') I(t') d(t') dt' \,.
\end{equation}
The beam is a series of short bunches:
\begin{equation}
	\label{eq:I_t}
	I(t) = \sum_{m=-\infty}^{\infty} I_0 t_b \delta_D(t - m t_b) \,,
\end{equation}
where $\delta_D$ is the Dirac delta-function.

\begin{figure}[h!]
	\includegraphics[width=86mm]{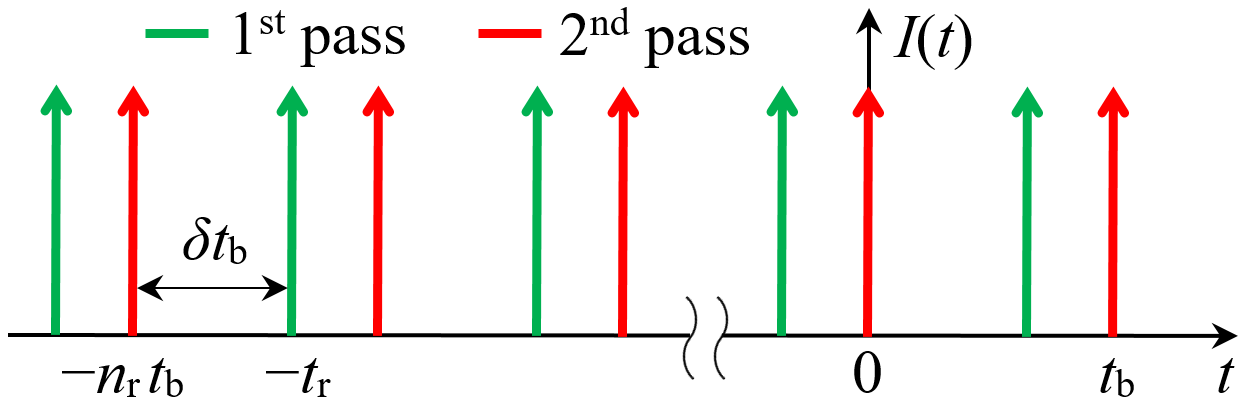}
	\caption{\label{fig:BeamCurrent}Beam current through the cavity (schematic view).}
\end{figure}

According to \cite{Hoffstaetter:2004BBU}, integer ($n_r$) and fractional ($\delta$) part of the ratio $t_r/t_b$ can be introduced as follows:
\begin{equation}
	t_r = (n_r - \delta) t_b \,, \quad 0 \leqslant \delta < 1 \,.
\end{equation}
Beam current through the cavity is schematically shown in Fig.~\ref{fig:BeamCurrent}.

For the MESA case all arcs should have $\delta=0$, except of the PIT arc which should have $\delta=\frac{1}{2}$. This provides steady acceleration of the bunches from the injector to the experimental area and steady deceleration of the used bunches on the way to the beam dump. However formulae for stability analysis can be obtained for arbitrary value of $\delta$.

Stability analysis implies searching for a harmonic solution for $V_{\lambda}(t)$ but it can not be a pure harmonic function $V_{0\lambda} e^{-i \omega t}$ because at least it is not smooth at $t = n t_b$ where $n$ is integer. As a workaround $V_{\lambda}(t)$ can be considered as a discrete function on a uniform grid with a period $t_b$. Samples should be taken at the time points $t_m = (m + \delta) t_b$ for integer $m$ (when bunches enter the cavity for the first time) because $V_{\lambda}(t_m)$ appear in the right-hand side of (\ref{eq:V_int}) after substitution of (\ref{eq:d_t}) and (\ref{eq:I_t}). Then $V_{\lambda}(t)$ contains harmonics with frequencies $\omega + \frac{2 \pi n}{t_b}$ and can be represented as Fourier series:
\begin{equation}
	V_{\lambda}(t) = \sum_{n=-\infty}^{\infty} V_n e^{-i \left( \omega + \frac{2 \pi n}{t_b} \right) t} \,.
\end{equation}
Fourier amplitudes $V_n$ are not important for the stability analysis because all harmonics are either stable or unstable depending on the sign of $\Im \omega$.

Substituting (\ref{eq:Wt_t}), (\ref{eq:d_t}) and (\ref{eq:I_t}) into (\ref{eq:V_int}) and taking into account that summation index should take integer values one obtains:
\begin{multline*}
	e^{-i \omega t} = \frac{K}{2 i} \sum_{m=-\infty}^{t/t_b - \delta} e^{-\Gamma_{\lambda} t} e^{\Gamma_{\lambda} m t_b} \times \\
	\times \left( e^{i \omega_{\lambda} t} e^{-i \omega_{\lambda} m t_b} - e^{- i \omega_{\lambda} t} e^{i \omega_{\lambda} m t_b} \right) e^{-i \omega m t_b} e^{i \omega t_r} \,,
\end{multline*}
where
\begin{equation}
	K = \frac{T_{12} q_e}{p c} \left(\frac{R}{Q}\right)_{\lambda} \frac{\omega_{\lambda}^2}{2 c} I_0 t_b = \frac{T_{12} q_e}{2 p} \left(\frac{R}{Q}\right)'_{\lambda} I_0 t_b \,.
\end{equation}
Here the only difference to \cite{Krafft:1990BBU} is presence of $\delta$ in the upper summation limit. After summation of geometric series this becomes:
\begin{multline}
	\label{eq:Dispersion1}
	1 = \frac{K}{2 i} e^{i \omega n_r t_b - \delta \Gamma_{\lambda} t_b} \times \\
	\times \left( \frac{e^{i \delta \omega_{\lambda} t_b}}{1 - e^{\left(-\Gamma_{\lambda} + i \omega + i \omega_{\lambda} \right) t_b}} - \frac{e^{-i \delta \omega_{\lambda} t_b}}{1 - e^{\left( -\Gamma_{\lambda} + i \omega - i \omega_{\lambda} \right) t_b}} \right)
\end{multline}
or
\begin{multline}
	\label{eq:Dispersion2}
	K = 2 e^{- i \omega n_r t_b + \delta \Gamma_{\lambda} t_b} \times \\
	\times \frac{\cos \left( \omega t_b + i \Gamma_{\lambda} \right) - \cos (\omega_{\lambda} t_b)}{e^{\Gamma_{\lambda} - i \omega t_b} \sin (\delta \omega_{\lambda} t_b) - \sin ((\delta - 1) \omega_{\lambda} t_b)}
\end{multline}
This is a dispersion relation connecting injector current and effective HOM voltage frequency. The result is exactly the same as in \cite{Hoffstaetter:2004BBU}.

\section{Beam Breakup Instability Theory}

Several generalizations are necessary:
\begin{enumerate}
	\item multiple recirculations;
	\item multiple cavities;
	\item multiple HOMs in each cavity;
	\item two transversal degrees of freedom.
\end{enumerate}

From now on the HOM set of the whole machine should be considered. If a HOM with the same properties is excited in two different cavities then it should be treated as two different HOMs. Therefore, HOMs should be numbered continuously throughout the machine. HOM numbering is not affected by the number of recirculations and the machine topology in general. Vice versa, transport matrices depend strongly on the topology but not on the HOM properties. Transport matrix from the $I$\nobreakdash-th pass of the $l$\nobreakdash-th cavity to the $J$\nobreakdash-th pass of the $m$\nobreakdash-th cavity may be denoted as $\mathbf{T}^{IJlm}$. Values $t_r$, $n_r$ and $\delta$ should be replaced with $t^{IJ}_{lm}$, $n^{IJ}_{lm}$ and $\delta^{IJ}_{lm}$ correspondingly. Also for convenience particle momentum at the $I$\nobreakdash-th pass of the $l$\nobreakdash-th cavity may be denoted as $p^I_l$.

HOM polarization should be taken into account to study bunch dynamics with two transversal degrees of freedom. Assume that dipole HOMs $\lambda$ and $\mu$ with polarization angles $\theta_{\lambda}$ and $\theta_{\mu}$ are excited in cavities at the beginning and at the end of a linear lattice region with transport matrix $\mathbf{T}^{IJ(lm)}$. At the entrance a bunch receives a transversal kick proportional to the distance to the first polarization axis in a direction perpendicular to this axis. At the exit this kick transforms into additional distance to the second polarization axis with the following proportionality coefficient:
\begin{multline*}
	\widehat{T}^{IJ}_{\lambda\mu} = \\
	= T_{12}^{IJl(\lambda)m(\mu)} \cos \theta_{\lambda} \cos \theta_{\mu} + T_{34}^{IJl(\lambda)m(\mu)} \sin \theta_{\lambda} \sin \theta_{\mu} + \\
	+ T_{32}^{IJl(\lambda)m(\mu)} \sin \theta_{\lambda} \cos \theta_{\mu} + T_{14}^{IJl(\lambda)m(\mu)} \cos \theta_{\lambda} \sin \theta_{\mu} \,.
\end{multline*}
Here the fact is explicitly indicated that the index of a HOM unambigiously defines the index of the cavity containing it. This expression should be used in (\ref{eq:d_t}) instead of $T_{12}$ (which corresponds to $\theta_{\lambda} = \theta_{\mu} = 0$).

If summation over HOMs from (\ref{eq:Wt_t}) is introduced into (\ref{eq:V_int}), then from (\ref{eq:Dispersion1}) follows that inverse injector current $\frac{1}{I_0}$ is an eigenvalue of $N_c N_m \times N_c N_m$ matrix $\mathbf{W} (\omega)$ with elements $W_{\mu \lambda} (\omega)$:
\begin{eqnarray*}
	W_{\mu \lambda} (\omega) \!=\! \frac{t_b q_e}{4 i} \!\sum_{I;J>I} \frac{\widehat{T}^{IJ}_{\lambda\mu}}{p^I_{l(\lambda)}} \!\left(\frac{R}{Q}\right)'_{\!\!\mu} \!e^{i \omega n^{IJ}_{l(\mu)m(\lambda)} t_b} \times \\
	\times \left( \frac{e^{(i \omega_{\lambda} + \Gamma_{\lambda}) \delta^{IJ}_{l(\mu)m(\lambda)} t_b}}{1 \!-\! e^{\left(-\Gamma_{\lambda} + i \omega + i \omega_{\lambda} \right) t_b}} \!-\! \frac{e^{(-i \omega_{\lambda} + \Gamma_{\lambda}) \delta^{IJ}_{l(\mu)m(\lambda)} t_b}}{1 \!-\! e^{\left( -\Gamma_{\lambda} + i \omega - i \omega_{\lambda} \right) t_b}} \right).
\end{eqnarray*}

\section{Complex Current Plot}

The eigenvalues of matrix $\mathbf{W} (\omega)$ can not be found analytically, thus indirect way is used. There are no infinitely growing solutions for $I_0 = 0$, this means that all $\omega$'s have negative imaginary parts. An instability emerges when $I_0$ exceeds a threshold value, this means that one of the $\omega$'s gets a positive imaginary part. Therefore, $\omega$ should be varied along the real axis within some region which spans the frequencies $\omega_{\lambda}$ of all excited HOMs. In general, the resulting current values will be complex, then the value with zero imaginary part and the smallest positive real part is the threshold current.

This leads to complex current plot technique when $I_0(\omega)$ values are plotted as a curve on the complex plane and then the closest to zero intersection of this curve with the positive real semiaxis is searched for. But in calculations this curve is represented as a set of points with varying step between them. Moreover, the number of curves is equal to the number of HOMs, and eigenvalue solver does not guarantee the order of the eigenvalues. This means that their order may be different for adjacent $\omega$ values. Without special countermeasures this would lead to jumps between curves and possible spurious intersections. The following algorithm was used to solve the problem: for each $I_0$ on the current step the closest value from the previous step is found (values with smaller $|I_0|$ are considered first), and values from the current step are sorted accordingly. Then the order of the curves is preserved and for sufficiently small step in $\omega$ intersection $I_*$ with real axis between adjacent values $I_1$ and $I_2$ can be found as follows:
\[
	I_* = \frac{\Re I_1 \Im I_2 - \Re I_2 \Im I_1}{\Im I_2 - \Im I_1} \,, \quad \textrm{if} \quad \Im I_1 \Im I_2 \leqslant 0 \,.
\]

\section{Parameter Spread}

This section is devoted to the estimation of different uncertainties which may affect the threshold current. Cavity manufacturing tolerances may lead to a HOM parameters spread. Beamline element misalignments and power supply errors may lead to a transport matrix elements spread.

\begin{figure}[h!]
	\includegraphics[width=86mm]{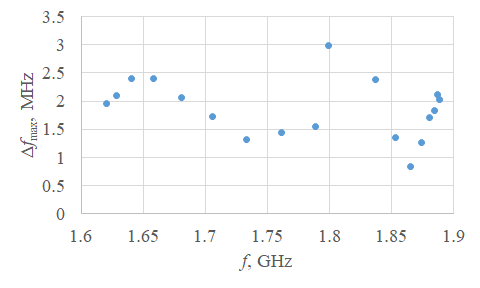}
	\caption{\label{fig:FrequencySpread}Measured frequency spread.}
\end{figure}

\begin{figure}[h!]
	\includegraphics[width=86mm]{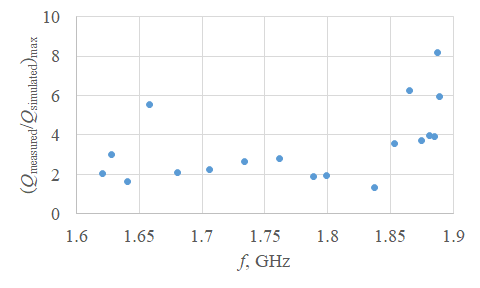}
	\caption{\label{fig:QfactorSpread}Maximal Q-factor ratio in measurements and simulations.}
\end{figure}

The frequencies and Q-factors of the first 36 dipole HOMs were measured twice for 4 MESA cavities (No. 7, 8, 9 and 10): during cold tests at DESY Hamburg and horizontal tests at HIM \cite{Stoll:2018dij}. The basic values of frequencies and Q-factors were taken from \cite{Ackermann:289143}. The maximal deviations of measured frequencies from the basic values are presented in Fig.~\ref{fig:FrequencySpread}. The maximal ratios of measured Q-factors and the basic values are presented in Fig.~\ref{fig:QfactorSpread}.

\begin{figure}[h!]
	\includegraphics[width=86mm]{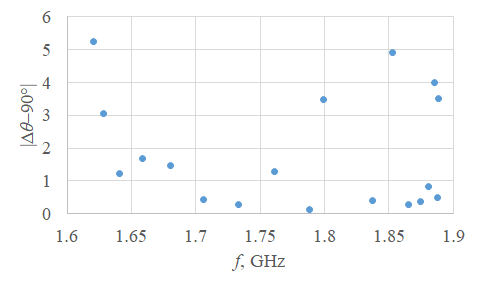}
	\caption{\label{fig:PolarizationSpread}Deviation from orthogonality in dipole HOM pairs.}
\end{figure}

In a cylindrically symmetric cavity the polarization angle of the first dipole HOM in a pair is arbitrary in the range $-\pi/2 \dots \pi/2$, the second is exactly orthogonal. In presence of high frequency power couplers and HOM couplers these angles have certain values, moreover the exact orthogonality is lost. However, production tolerances may randomize polarization angles back but deviation from orthogonality may remain. Shunt impedances and polarization angles were calculated with CST MICROWAVE STUDIO. Deviations from orthogonality within HOM pairs are presented in Fig.~\ref{fig:PolarizationSpread}.


Quadrupole gradient and quadrupole roll errors may lead to matrix elements variation. For each configuration separate lattice file with random error values is generated, then the matrix elements are calculated using \verb'elegant' \cite{Borland:2000gvh}. General tolerance to quadrupole gradient errors should be taken as a value of $\Delta K/K$, usually it does not exceed $10^{-3}$. General tolerance to quadrupole roll errors usually is not more than 1~mrad.

Other lattice related errors do not affect the transport matrix elements significantly because the optics is linear. Reference orbit offsets and cavity misalignments may only change the equilibrium HOM voltage (if any) but not its frequency and, therefore, can not affect threshold current. It was shown in \cite{Hoffstaetter:2004BBU} that the beam deflection caused by HOMs may lead to large reference orbit distortions but this happens at much higher current values than the BBU threshold current.

\section{Worst Case Configuration}

Assume that each HOM interacts only with itself on different passes but not with other HOMs (of the same type but excited in different cavities and with slightly different parameters due to cavity manufacturing tolerances). Therefore, for each HOM the worst combination of cavity manufacturing tolerances which leads to minimal possible threshold current can be found.

Introduce in (\ref{eq:Dispersion2}) the following notation:
\[
	\epsilon = \Gamma_{\lambda} t_b \,,	\quad	\omega = \omega_{\lambda} + \Delta \omega \,.
\]
If $\Delta \omega t_b \ll 1$, $\epsilon \ll 1$ which is usually the case, then (\ref{eq:Dispersion2}) leads to
\begin{eqnarray*}
	K \approx -2 (\Delta \omega t_b \cos(\omega t_r) + \epsilon \sin(\omega t_r) + \\
	+ i (\epsilon \cos(\omega t_r) - \Delta \omega t_b \sin(\omega t_r))) \,.
\end{eqnarray*}
$K$ is real, therefore 
\[
	\Delta \omega t_b \approx \epsilon \, \mathrm{ctg}(\omega t_r)
\]
and
\[
	K \approx -\frac{2 \epsilon}{\sin(\omega t_r)} \,.
\]

For one cavity with one HOM, multiple recirculations and two transversal degrees of freedom this leads to:
\begin{equation}
	\label{eq:dispersionsimplifiedcomplicated}
	I_{th}^{\lambda} \!\approx\! -\frac{2 c^2}{q_e \!\left(\frac{R}{Q}\right)_{\lambda} \!\!Q_{\lambda} \omega_{\lambda}} \frac{1}{\displaystyle \sum_{I;J>I} \frac{\widehat{T}^{IJ}_{\lambda\lambda}}{p^I_{l(\lambda)}} \sin\!\left(\!(\omega_{\lambda} \!+\! \Delta \omega) t^{IJ}_{l(\lambda)l(\lambda)}\!\right)} .
\end{equation}

The Q-factor and the shunt inpedance appear only in a common factor, thus their effect on the threshold current is straightforward. So, the maximal possible Q-factor value should be chosen and the denominator of the second fraction in (\ref{eq:dispersionsimplifiedcomplicated}) should be maximized by absolute value over variables $\Delta \omega$ and $\theta_{\lambda}$.

The matrix elements $T_{14}$ and $T_{32}$ are always small because MESA has no transversal coupling, thus all extrema are at $\theta_{\lambda} = 0$ and $\theta_{\lambda} = \pi/2$. Then only $\Delta \omega$ is to be varied. The relative HOM polarization is not important because interaction between HOMs is neglected. Therefore, two configurations for each HOM pair should be checked: the first HOM is horizontally polarized, the second is vertically polarized and vice versa. The one with smaller threshold current is the worst case configuration.

\section{Simulation Results}

\begin{figure}[h!]
	\includegraphics[width=86mm]{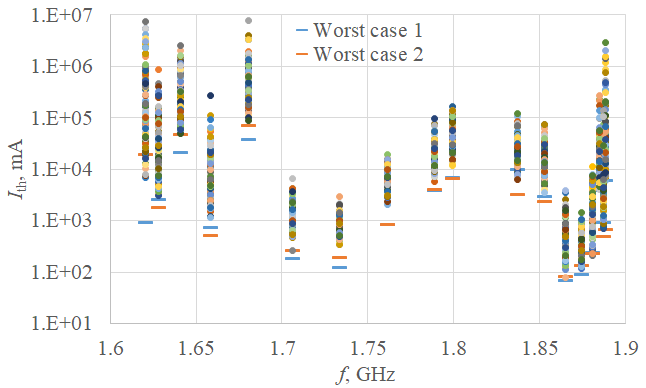}
	\caption{\label{fig:ThresholdCurrent}Threshold current for 100 random machine configurations and 2 worst case configurations.}
\end{figure}

In Fig.~\ref{fig:ThresholdCurrent} threshold currents for 100 different random machine configurations are shown, with two worst case configurations added. The frequency, Q-factor and deviation from orthogonality spread for each HOM pair is chosen in accordance to the maximal spread obtained in measurements, HOM polarization angle is random, quadrupole gradient relative error is $10^{-3}$, quadrupole roll error is 1~mrad. The technique for predicting the worst case configuration proposed in the present paper seems to be quite reliable.

\begin{figure}[h!]
	\includegraphics[width=86mm]{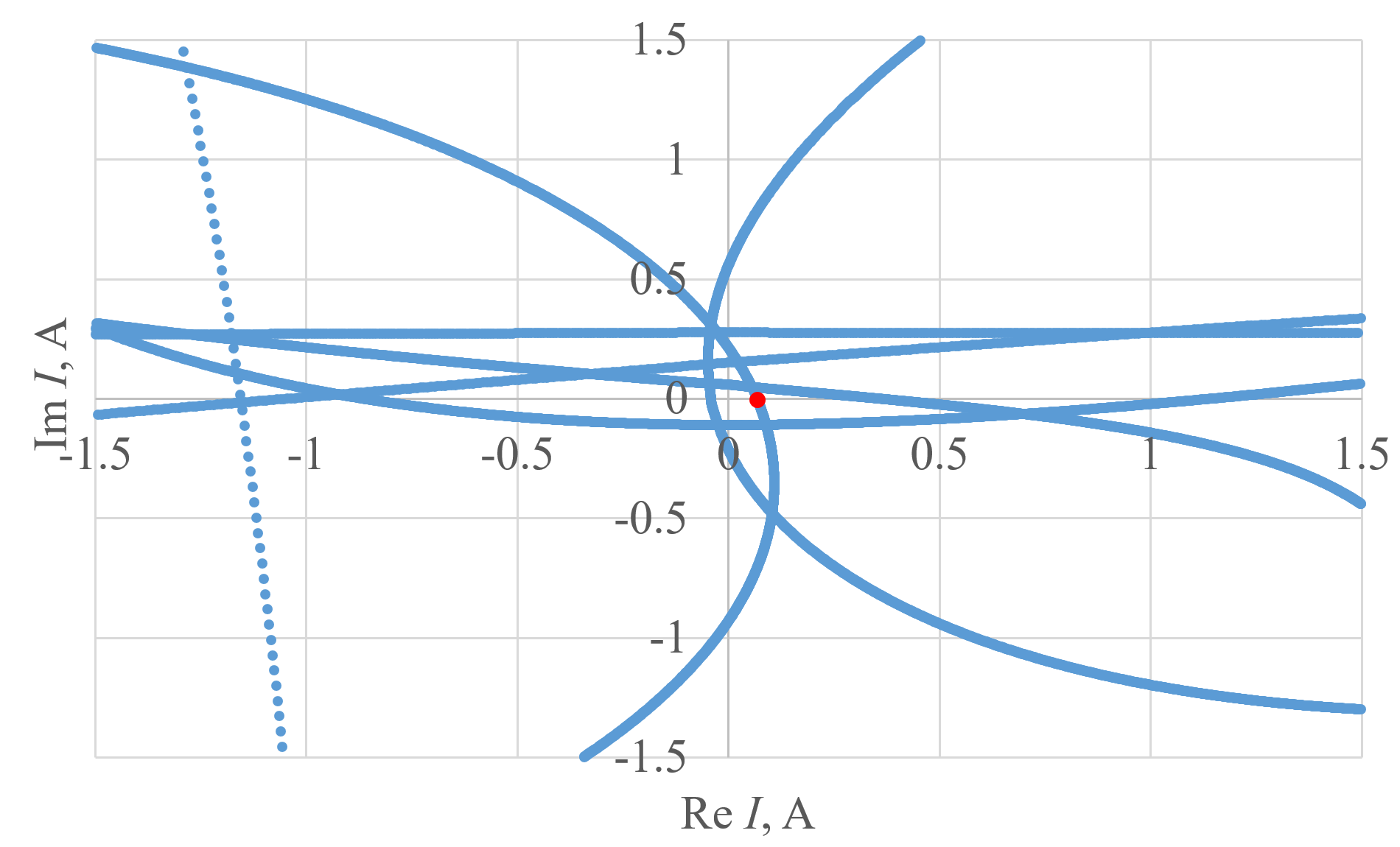}
	\caption{\label{fig:ComplexCurrentPlot}Complex current plot for minimal threshold current configuration. The threshold current value is marked with a red point.}
\end{figure}

\begin{figure}[h!]
	\includegraphics[width=86mm]{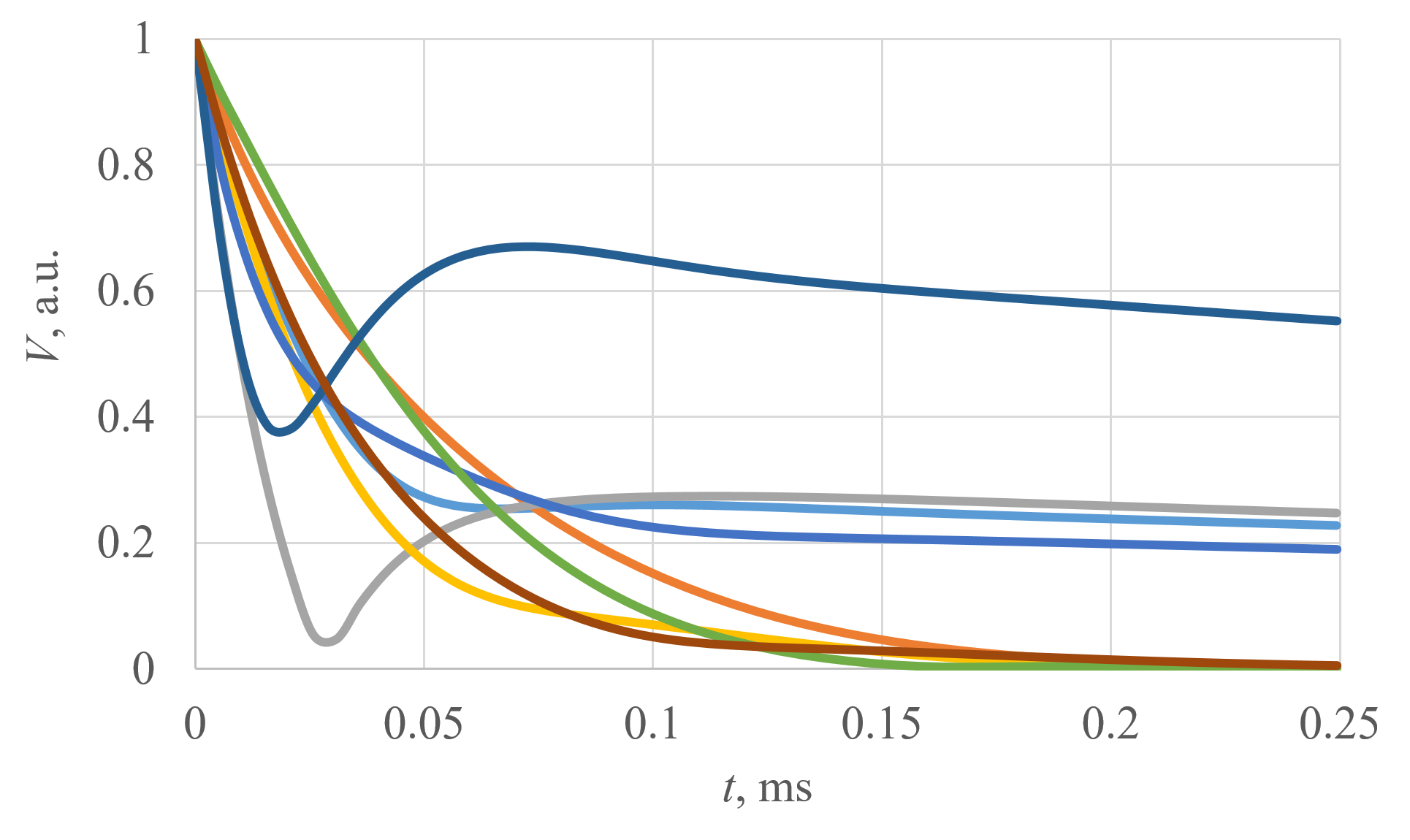}
	\caption{\label{fig:VoltageStable}HOM voltages time dependence for a current below threshold.}
\end{figure}

\begin{figure}[h!]
	\includegraphics[width=86mm]{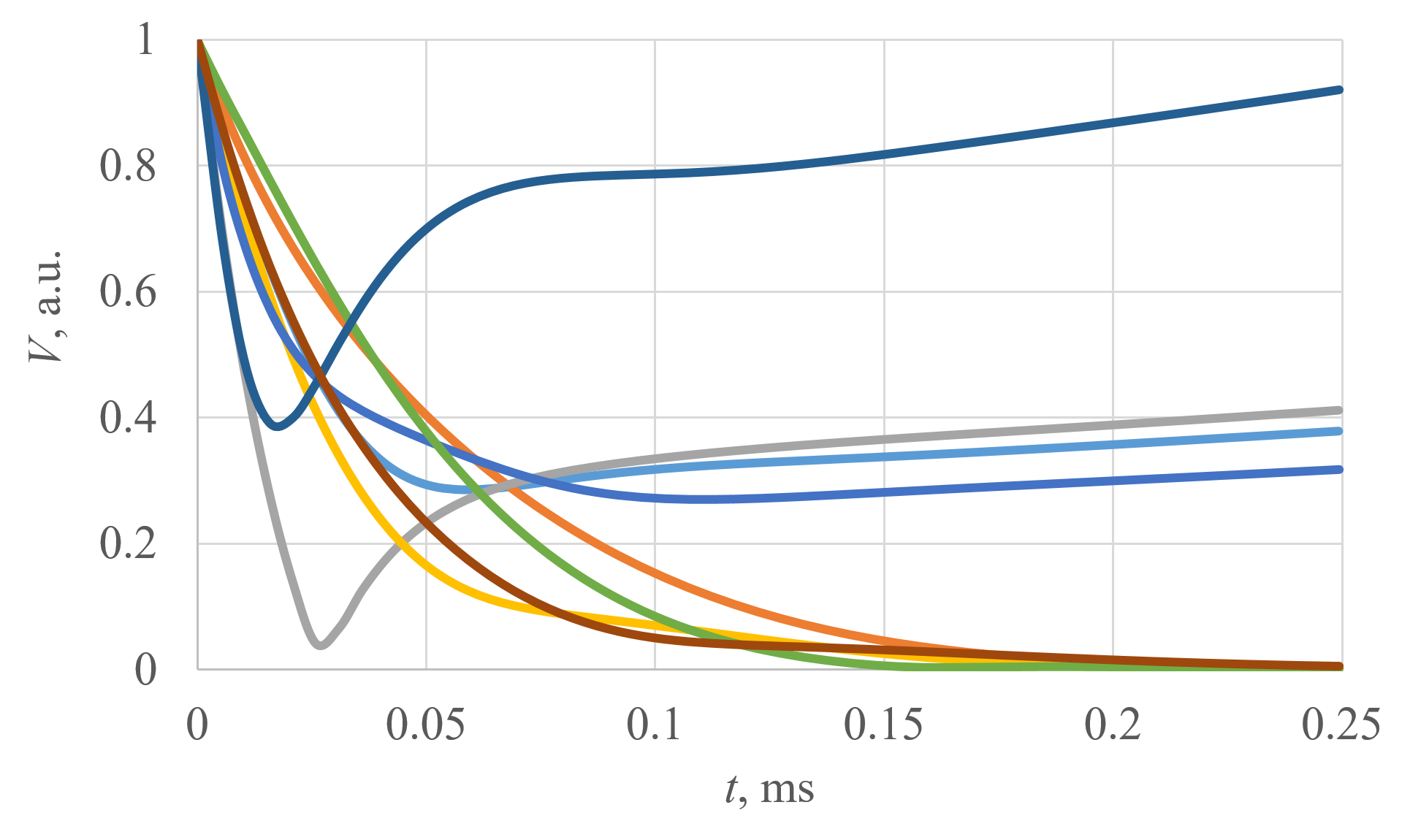}
	\caption{\label{fig:VoltageUnstable}HOM voltages time dependence for a current above threshold.}
\end{figure}

The minimal threshold current $I_{\textrm{th}} = 68.28 \, \textrm{mA}$ was obtained for the worst case configuration with HOMs No. 39 and 40 at 1.87~GHz. Figure~\ref{fig:ComplexCurrentPlot} shows the central part of complex current plot for this particular case, the threshold current value is marked with a red point. Current values of $I = 67 \, \textrm{mA}$ and $I = 70 \, \textrm{mA}$ were chosen to demonstrate stable and unstable motion. For both values the HOM voltages in all cavities over time are shown in Fig.~\ref{fig:VoltageStable} and \ref{fig:VoltageUnstable}. It can be seen that after a relatively short transition process HOM voltages begin to constantly fall (below threshold) or rise (above threshold).

This is an example of a visual check performed for several tens of machine configurations to make sure that the results obtained with both techniques are consistent. In principle, the threshold current may be found using only bunch tracking with varying current (e.g. with bisection method, as in the \verb'bi' code by I.~Bazarov \cite{Bazarov:bi}) but care should be taken because the exponential fit of the voltages or final coordinates may give different decrement signs depending on the considered time interval.


\section{Conclusion}

A set of scripts in Python were developed for calculation of the BBU instability threshold current using two different techniques: a complex current plot technique (based on the well-known BBU instability theory) and direct bunch tracking. The results obtained with both techniques are consistent for various dipole HOM and lattice configurations in MESA. A method proposed for predicting the worst case configuration with the smallest threshold current works also well. With the most dangerous uncertainties taken into account, the threshold current is not less than 68~mA which is well above the 10~mA design current for the second stage of MESA operation in ER mode. Therefore no additional measures need to be taken to avoid negative consequences of the BBU effect.

\begin{acknowledgments}
The authors thank W.~M{\"u}ller and W.~Ackermann for consultations in TESLA cavity and CST MICROWAVE STUDIO related topics, A.~Khan and D.~Simon for MESA lattice files, H.~De~Gersem for assistance in preparation of the present paper.
\end{acknowledgments}

\bibliography{BBU}

\end{document}